# Ultra-clean isotope engineered double-walled carbon nanotubes as tailored hosts to trace the growth of carbyne

Weili Cui, *[a] Ferenc Simon, [b] Yifan Zhang, [c, d] Lei Shi, *[c] Paola Ayala [a] and Thomas Pichler*[a]

| | |
|---|---|
| [a] | Dr. W. Cui, * Prof. P. Ayala, Prof. T. Pichler * <br> Faculty of Physics <br> University of Vienna <br> Boltzmanngasses 5, Vienna, 1090, Austria <br> E-mail: weili.cui@univie.ac.at, thomas.pichler@univie.ac.at |
| [b] | Prof. F. Simon <br> Department of Physics <br> Institute of Physics, Budapest University of Technology and Economics <br> Muegyetem rkp. 3., H-1111 Budapest, Hungary |
| [c] | Dr. Y. Zhang, Prof. L. Shi * <br> State Key Laboratory of Optoelectronic Materials and Technologies, Nanotechnology Research Centre, Guangzhou Key Laboratory of Flexible Electronic Materials and Wearable Devices, School of Materials Science and Engineering <br> Sun Yat-sen University <br> Guangzhou 510275, P. R. China <br> E-mail: shilei26@mail.sysu.edu.cn |
| [d] | Prof. Y. Zhang <br> School of Engineering Department <br> Huzhou University Institution <br> Huzhou, Zhejiang 313000, P. R. China |

Supporting information for this article is given via a link at the end of the document.

**Abstract:** Increasing attention is currently given to carbyne, the sp[1] hybridized one-dimensional carbon allotrope, because of its predicted outstanding mechanical, optical, and electrical properties. Although recently substantial progress has been reported on confined carbyne synthesized inside double-walled carbon nanotubes (DWCNTs), its formation mechanism and precursors for growth remain elusive. Here, we show a rational design of isotope engineered ultra-clean DWCNTs as tailored hosts to trace the growth of carbyne, which allows to identify the precursor and unravel the formation mechanism of carbyne during high-vacuum annealing at high-temperatures. Using this approach, ultra-clean DWCNTs with 80.4% $^{13}$C-enriched inner walls and outer tubes of naturally abundant served to unambiguously prove that only the carbonaceous materials inside the DWCNTs can act as precursors. The exchange of C atoms between inner and outer tubes happens without any growth of carbyne. After applying a secondary oxidation step, it is possible to produce the carbonaceous precursors from the partially oxidized DWCNTs. In this manner, not only carbyne with a record of ~28.8% $^{13}$C enrichment is grown, but concomitant healing, reorganization and regrowth of the DWCNTs occurs. This work enables to identify the precursor and trace the growth mechanism of confined carbyne with engineered properties. This is a crucial step, towards accessing the full application potential of confined carbyne hybrids by tailoring not only the isotopic fillers, but also the inner and outer tubes of the DWCNT hosts.

## Introduction

Carbyne is an allotrope of carbon with sp[1] hybridization, which renders excellent mechanical, optical, and electrical properties.[1–8] Its synthesis was first achieved from the polyyne. However, the longest polyyne reported to date only consists of 48 carbon atoms and its properties rely on its length.[9–12] Only recently, the materialization of extreme long linear carbon chains (LCCs) was achieved using double-walled carbon nanotubes (DWCNTs) as templates and precursors. [13] These long encapsulated LCCs were the first to show carbyne properties that depended on the diameter of the hosting nanotubes rather than on the length of the carbon chain.[14,15] A considerable amount of research has been devoted to increasing the growth ratio of such confined carbyne (CC) as well as to investigating its giant Raman cross section and anti-Stokes Raman spectral response.[16,17]

Regarding the growth process, it has been *in-situ* monitored by applying laser annealing of carbon nanotubes (CNTs) locally. [18] However, the carbon source for the CC is still an open question, since it has not been possible to disambiguate whether the carbon residues already existing inside the carbon nanotubes before the synthesis or carbon nanotube walls themselves are responsible for providing the carbon atoms that form the CC. Therefore, understanding the formation mechanism is of fundamental interest to ultimately optimize and increase the formation yield. This is required for the CC future applications as a semiconductor [19] in quantum spin transport, nanoelectronics, and nuclear magnetic resonance, etc. [3]

The first step on the synthesis of carbyne corresponds to the fabrication of polyyne, which are short polymerized one-dimensional chains that were synthesized by a direct organic chemistry bottom-up approach using end-capping chemical groups to stabilize them. However, the length of the chains was extremely limited.[11,20] Still, long LCCs can be achieved by coalescing short polyynes inside CNTs at high temperature, but the chain length highly depends on the filling level with polyynes. Since the encapsulation of polyyne is very challenging, increasing the chain length through this pathway was extreme difficult.[10,21,22] Numerous common approaches to grow LCCs have appeared using multi-walled CNTs [23] DWCNTs [13] and single-wall CNTs (SWCNTs) [24] as hosts produced by arc-discharge or CVD methods. Experiments with such nanotubes can yield CC with more than 6000 carbon atoms. [13] However, those results correspond to experiments using the residual carbon existing inside the CNT, proving that the yield of CC is not high given a limited carbon source. [13,15,23] Our recent work showed that introducing additional carbon sources from liquid precursors (e. g., methanol) allow to enhance the yield of CC by a factor of 2. [25] Furthermore, isotope labelling of CC was made feasible using liquid precursors, which were proven as external carbon source. However, it is still an open question whether the carbon atoms in the CNTs contribute to the carbon sources or not.

To address these open issues, here we utilized $^{13}$C labelled DWCNTs as tailored hosts to trace the growth process of the CC. Natural and isotopic C$_{60}$ fullerenes were filled inside SWCNTs with average diameter of 1.4 nm. This resulted into peapod structures, which were transformed later into DWCNTs with natural (1.1% $^{13}$C) and isotopic (80.4% $^{13}$C) inner tubes, thus labeled as $^{Nat}$DWCNTs and $^{Iso}$DWCNTs, respectively. The as-prepared DWCNTs were then used to synthesize the CC by applying oxidation in air followed by high-temperature annealing in vacuum. The directly annealed samples without oxidation were also studied for reference. We found that the CC forms when both the oxidation and annealing were applied and the CC has a highest 28.8 % $^{13}$C content up to now. The oxidation not only creates defects on the walls of the DWCNTs, but it also produces precursors from partially oxidized DWCNTs, which further contributes to the formation of the CC and the reorganization of the DWCNTs. Our results highlight how isotope-engineered DWCNTs, as tailored hosts, enable to trace the growth of the CC by tracing their $^{13}$C contents, since the isotope-related shifts in the corresponding Raman modes allow to separate the response from the inner and outer tubes. This enables us to analyze the changes in isotope concentration to reveal the formation of the CC and the reorganization of DWCNTs from isotopic precursors. Furthermore, it has been possible to experimentally prove the theoretically predicted exchange of carbon atoms between inner and outer tubes.[26,27] This work also enables us to unravel the growth of CC and the inner tubes from partially oxidized inner and outer tubes worked as precursors. This implies the clear possibility to trace and tailor the synthesis of CC by controlling not only the fillers, but also the hosting carbon nanotubes.

## Results and Discussion

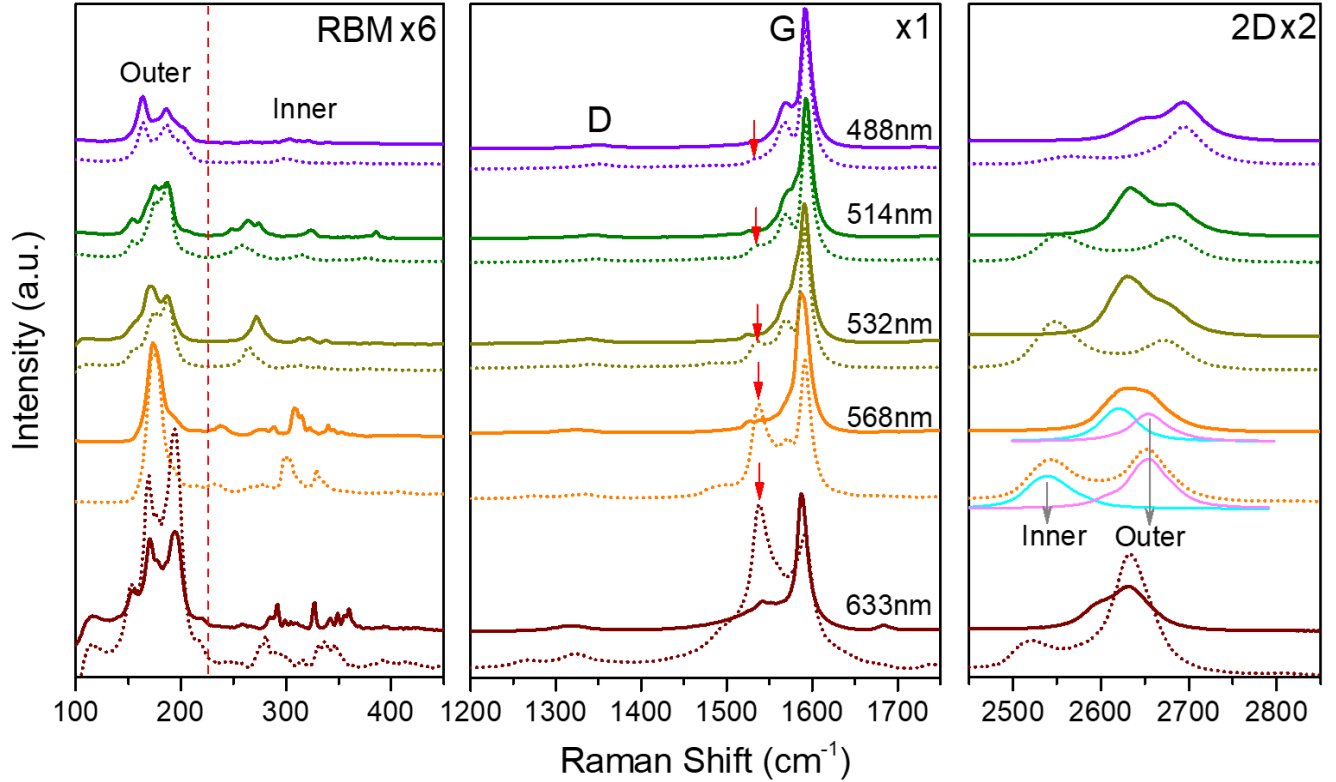

**Figure 1.** Raman spectra of the $^{Nat}$DWCNTs (solid lines) and $^{Iso}$DWCNTs (dotted lines) excited by lasers with different wavelengths, the downshifted G-mode originated from the $^{13}$C labelling were marked by the red arrows.

The initial $^{Nat}$DWCNTs and $^{Iso}$DWCNTs were prepared from the natural and isotope labelled C$_{60}$ (~80%) filled into arc-discharge SWCNTs with a narrow diameter distribution (1.4 ± 0.2 nm) via peapod annealing at 1250 °C in high vacuum. [28] Their Raman spectra are shown in **Figure 1**. The radial breathing mode (RBM) in the Raman spectra of the CNTs is associated with the atoms around the circumference vibrating in the radial direction, which is located in the Raman frequency range of 100-400 cm$^{-1}$ and it is inversely related to the CNT diameter. [29] The RBM peaks above 230 cm$^{-1}$ correspond to the thin inner tubes with a diameter that be calculated by the formula $D = 234/(\omega_{RBM} - 10)$, where D is the diameter of the tubes in nm and $\omega_{RBM}$ is the Raman frequency of the RBM peaks in cm$^{-1}$. [30] With different resonance excitations, the Raman signals of the inner tubes in the RBM region mainly appear from 240 to 370 cm$^{-1}$, thus, the calculated diameters of the inner tubes are 0.65-1.02 nm. An overall downshift of these RBM peaks was observed for the $^{Iso}$DWCNTs, which clearly shows that isotope labelling is achieved for the inner tubes. In addition, downshifts were observed for the G-mode (marked as the red arrows in **Figure 1**) and the 2D-mode of the inner tubes. After fitting the 2D-band (**Figure S1**), we found that the components of the outer tubes remain at the same frequency, whereas the components corresponding to the inner tubes downshift significantly. This can be attributed to the isotope effect on the inner tubes. The downshift of the Raman frequency from the increased mass of $^{13}C$ atoms with concentration $c$ can be calculated by $(\nu_0 - \nu)/\nu_0 = 1 - \sqrt{(12+c_0)/(12+c)}$, where $\nu_0$ or $\nu$ are the frequencies with or without $^{13}C$ isotope, respectively, whereas $c_0 = 0.011$ represents the natural abundance of $^{13}C$ in carbon materials. [28] Therefore, the relative shifts of the 2D modes excited with different laser wavelengths can be used as reference to calculate the $^{13}C$ enrichment. This results into an average value $c$ of 80.4% for the inner tubes of the $^{Iso}$DWCNTs (**Table S1**), which is consistent to the isotope ratio of the $C_{60}$ used for filling.

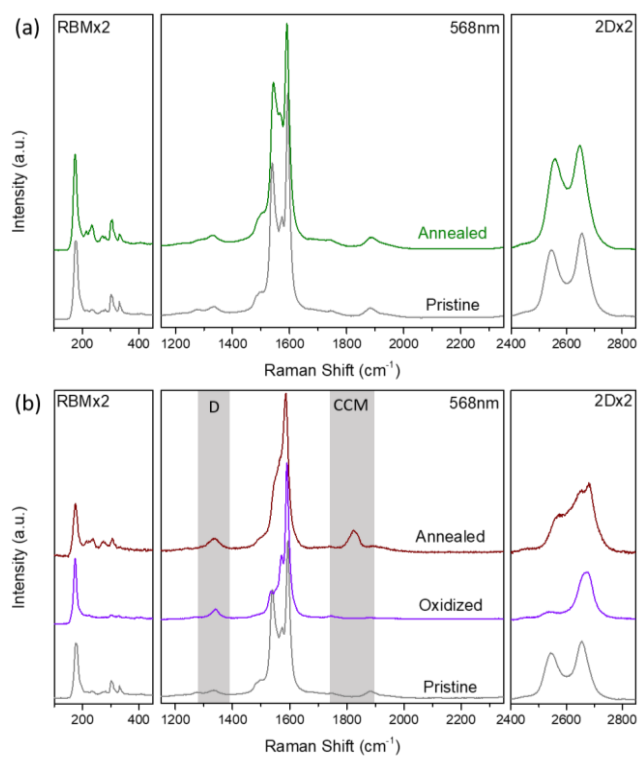

**Figure 2.** Raman spectra of the one-step and stepwise annealed $^{Iso}$DWCNTs excited at 568nm. (a) The one-step annealed $^{Iso}$DWCNTs with no obvious change compared to that of the pristine reveals that the structure of the $^{Iso}$DWCNTs still remains intact. (b) The oxidized $^{Iso}$DWCNTs with increased intensity of the D-band suggests a defective nanostructure, which were further annealed and finally formed the isotope labelled confined carbyne as indicated by the newly appearing CCM signal.

In order to grow the CC, the $^{Nat}$DWCNTs (as-prepared) and $^{Iso}$DWCNTs were simultaneously annealed at 1500 °C in high vacuum. Previously, we applied the same method to transform residual carbon inside nanotubes [15] or organic solvent molecules[25] as precursors to synthesize the CC. However, no CC was formed when using the isotopic DWCNTs as precursor by a direct high-temperature annealing (marked as one-step annealing). This hints that the starting material is clean and no additional carbon atoms contribute to the formation of the CC (see **Figure 2a** and **Figures S2-S4**). According to our previous study on the improved oxidation stability of the thin inner tubes with the protection from the host SWCNTs, [31] a stepwise method was applied to prepare defective outer and inner tubes for the growth of the CC. The DWCNTs were first oxidized in air at 520 °C to reopen the tubes and to introduce defects on both the outer and inner tubes, which then act as carbon sources in a subsequent step to facilitate the formation of the CC during the annealing at 1500 °C for reconstruction of the DWCNT and formation of the CC. As shown in **Figure 2b**, the Raman spectra of oxidized $^{Iso}$DWCNTs show larger D/G ratios, which hint the successful introduction of defects. In addition, the RBM intensity of the inner tubes decreases significantly after the oxidation, confirming the damage of the inner tubes. This can also be observed by the simultaneously decreased intensity of the 2D-band, which can be associated to the inner tubes. This higher affinity of the inner tube for oxidation is due to the larger curvature of the inner tubes compared to that of the outer ones. In turn, this means that the inner shell deviates more from the stable $sp^2$ graphitic configuration, which results into an enhanced reactivity to oxygen. The Raman response of the oxidized sample after annealing shows a mode located at 1790-1860 $cm^{-1}$, as the confined carbyne mode (CCM) shown in **Figure 2b**, which demonstrates that the CC was successfully synthesized. [32] This illustrates how oxidation plays an important role in the growth of the CC by introducing defects on the CNT walls and producing carbonaceous precursors from the partially oxidized CNTs.

Since the optical band gap of the CC can be tuned by the intrinsic length of the chains as well as the environmental interactions with the hosting tubes, [14,19,32] a resonance Raman study was subsequently performed (as shown in **Figure S2-S4**) to evaluate different lengths of the CC synthesized inside different CNTs using various excitation laser wavelengths. Interestingly, the CC mode of the $^{Iso}$DWCNTs shows a significant broadening and downshift compared with that of the $^{Nat}$DWCNTs (**Figure 3d** and **Figure S5**), which demonstrates a high level of $^{13}C$ enrichment in the CC synthesized from the $^{Iso}$DWCNTs with random isotope distribution results in an inhomogeneous broadening of the Raman modes. [28]

In order to analyze the average $^{13}C$ ratios in the nanotubes after annealing, selective laser excitations were applied for chosen CNTs with different chirality. The laser choice also took into account the different band gaps, which could be potentially seen for the CC. As depicted in **Figure 3** and **Figure S6-S7**, the 2D-band of the outer tubes and inner tubes as well as the CCM of the CC were analyzed to obtain their isotope ratios. For the one-step annealed $^{Iso}$DWCNTs, no CC was formed (**Figure 2a** and **Figures S2-4**). However, the $^{13}C$ ratio of the inner tubes decreased from 80.4% to 68.5% while the $^{13}C$ ratio of the outer tubes increased from 1.1% to 4.1%, revealing that the carbon atoms were exchanged between the inner tubes and the outer tubes during the high temperature annealing process, most likely around the defects. This is a quantitative confirmation of the previous theoretical predictions of such effect. [26,27] In terms of the stepwise oxidized and annealed DWCNT samples (**Figure 3c&3d** and **Table 1**), the CC with a 28.8% $^{13}C$ ratio (the highest record up to now) were grown, revealing that the oxidation plays a crucial role for proving the isotopic precursors from partially oxidized CNTs. The $^{13}C$ ratio of the inner tubes further decreased to 57.4% besides the exchange effect (68.5%) and meanwhile the $^{13}C$ ratio of the outer tubes further increased to 6.5% besides the exchange effect (4.1%), since the isotopic precursors also contributed to the healing of defects and reorganization of tubes in addition to the carbon atoms exchange occurring with the unoxidized DWCNTs.

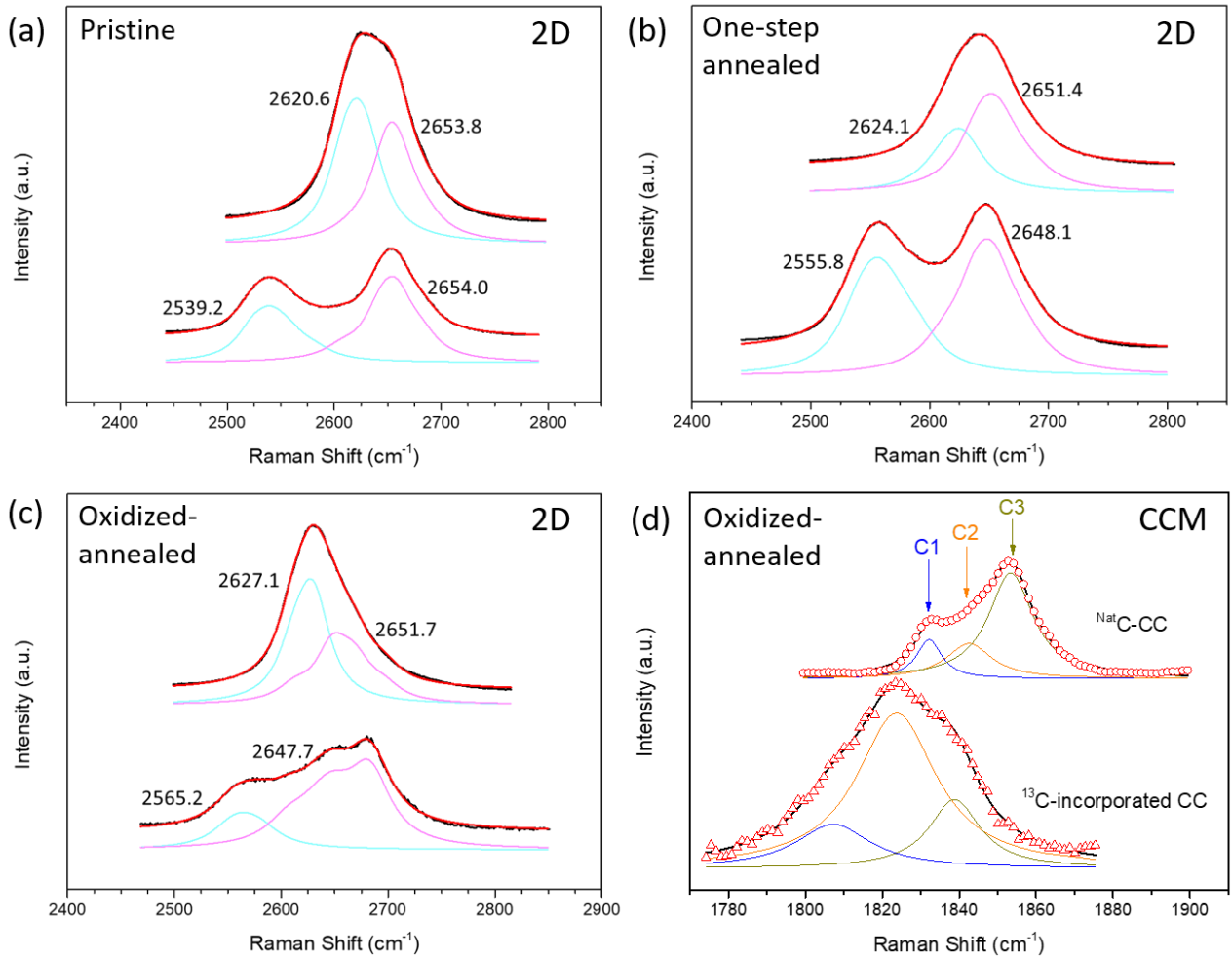

**Figure 3.** Line-shape analysis of the (a) pristine, (b) one-step annealed (c, d) stepwise oxidized and annealed [Nat]DWCNTs (upper) and [Iso]DWCNTs (lower) samples excited at 568 nm. The 2D-band of the outer tubes includes several components considering the diameter distribution. (d) The CCM of the stepwise oxidized and annealed samples were fitted with three Lorentzian peaks corresponding to the CCs with different band gaps.

**Table 1.** The calculated average $^{13}$C concentration of the outer tubes, inner tubes and confined carbyne.

| $^{13}$C concentration | Outer wall | Inner wall | CC |
|---|---|---|---|
| Pristine [Iso]DWCNTs | 1.1 % [28] | 80.4 ± 0.8 % | - |
| One-step annealed [Iso]DWCNTs | 4.1 ± 0.4% | 68.5 ± 2.2% | - |
| Stepwise oxidized and annealed [Iso]DWCNTs | 6.5 ± 0.8% | 57.4 ± 3.6% | 28.8 ± 1.4 % |

Based on our findings, we can propose the following growth model for confined carbyne as sketched in **Figure 4**. First, the isotope enriched inner walls with 80.4% $^{13}$C ratio inside the natural SWCNTs are obtained by encapsulating and then transforming the isotopic fullerenes at high temperature annealing in vacuum. Then, the DWCNTs with isotopic inner walls were treated via annealing either in one step or stepwise alternating oxidation and annealing. Since the $C_{60}$ with diameter of 0.7 nm allows to be transformed into an inner tube with the diameter of 0.7 nm and with the length of 0.7 nm, the maximum length of the inner tubes transformed from 100% filled peapods is around 70% of the outer tubes by considering the van der Waals distance of 0.35 nm between the $C_{60}$, yielding ~2.8 times carbon atoms in the outer as compared to the inner tubes. [33,34]

In the one-step annealed samples, no CC was formed, because no additional carbonaceous precursors were present to initiate the growth, which highlights that the starting DWCNTs are very clean, i.e., almost no carbon residues in the nanotubes. But the isotope ratios of the inner/outer tubes decreased/increased from 80.4%/1.1% to 68.5%/4.1%, respectively, suggesting the carbon atoms exchange between the inner and outer tubes during the high temperature annealing process. Considering 100% filling of the $C_{60}$ inside a SWCNT fully transformed inner tube with around 70% of the length of the SWCNTs, in our case since the isotope ratio changes of the inner and outer tubes are -11.9% and +3.0%, respectively, this indicates that the inner tubes actually only fill around 50% of the outer tubes. This means that the filling ratio of the $C_{60}$ in the peapods was not 100% but 70%. [34,35] Therefore, the outer tubes should accordingly contain 4 times amount of carbon atoms compared to that of the inner tubes.

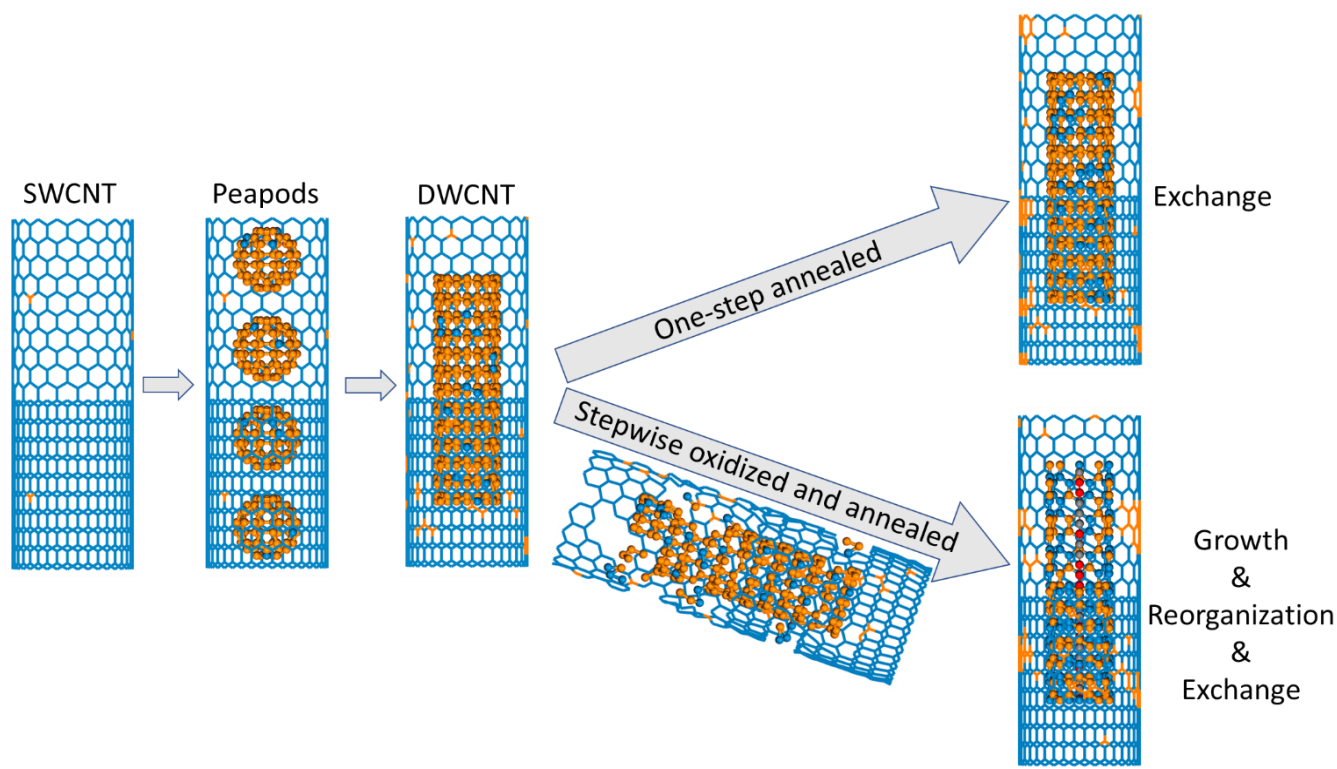

**Figure 4.** Schematic synthesis process of the growth of isotope labelled LCCs@DWCNTs. The blue and gray represent $^{12}$C content while the orange and red represent $^{13}$C content, respectively.

For the stepwise annealing assisted by oxidation, the DWCNTs with isotope enriched inner walls are first etched by oxygen to form defects on the walls of the DWCNTs, where small pieces from both the outer tubes with less $^{13}$C carbon atoms and inner tubes with enriched $^{13}$C carbon atoms were produced and then acted as precursors for the synthesis of the CC in the following annealing step. This yielded the CCs with an average $^{13}$C ratio of 28.8%. With the knowledge that the thin diameter inner tubes are less stable than the outer tubes, this observation allows us to estimate the oxidation degree of inner and outer tubes. In principle, if the carbon atom exchange achieves equilibrium, the $^{13}$C ratio of the CC should be (80.4%*50% +200%*1.1%) / (50%+200%) = 17%, which is much less than the experimental result. Such a difference in the isotope enrichment level is due to the different oxidation degree of the inner tubes and the outer tubes, which causes that the inner tubes supply more precursors for the formation of the CC. Therefore, it can be concluded that if the oxidation degree of the inner tube is around 2 times than that of the outer tubes, the $^{13}$C enrichment of the CC as can be calculated as (80.4%*100%+200%*1.1%)/3 = 27.5%, which is in line with our experimental value of 28.8%. Since the inner tubes transformed from the $C_{60}$ molecules are very defective in the beginning (due to the strongly non-sp$^2$ character of the carbon-carbon bonds) and with thin diameter, then, the inner tubes are easier to be oxidized than the outer tubes with much less defects. In addition, from the one-step annealed samples to the stepwise treated samples, we observed a further decrease of the $^{13}$C content in the inner tubes from 68.5% to 57.4% and an increase of the $^{13}$C content in the outer tubes from 4.1% to 6.5%. This can be fully explained by the reorganization of the defective tubes happens simultaneously with the CC's formation, which means that the atom exchange between the inner and outer tubes increases with the oxidation level because of the larger number of defects and the presence of additional $^{13}$C isotope precursors.

## Conclusion

We have successfully utilized the isotope engineered DWCNTs as tailored hosts to trace the growth of carbyne. The isotope related shifts of the Raman modes allow to separate the response of inner and outer tubes while tracing the $^{13}$C content in this system. This enables identifying the precursors of confined carbyne released by the partially oxidized DWCNTs along with the reconstruction of the defective nanotubes, as well as proving experimentally the exchange of carbon atoms between the inner and the outer tubes. Such ultra-clean DWCNTs cannot form CC due to no additional carbon source. However, they can act as a clean and well-defined host for the CC synthesis when an oxidation step is included. Since the oxidation process not only introduces defects on the walls in DWCNTs, but also produces small fragments of $^{13}$C enriched nanotubes. These can further act as precursors for the CC formation as well as reorganization of the DWCNTs through annealing. As a result, the CC with ~28.8% $^{13}$C ratio were finally obtained by using the partially oxidized CNTs as precursor, which is the highest isotope ratio achieved up to now. This work paves the way to access the full application potential of these engineered materials, being able to utilize defined precursors not only from the fillers but also controlling the defects of the carbon nanotubes.

## Experimental Section

*Synthesis of natural and isotope labelled confined carbyne*: SWCNTs with an average diameter of 1.40 ± 0.10 nm were made with the arc-discharge method. DWCNTs were prepared by vapor filling the $C_{60}$ molecules at 650 °C and then transforming them into inner walls inside the SWCNTs at 1250 °C. By choosing the natural or isotopic $C_{60}$, natural abundant (1.1%) or isotope labelled (around 80%) inner tubes can be synthesized on demand as reported previously, [28] and the obtained DWCNTs were thus named as $^{Nat}$DWCNTs or $^{Iso}$DWCNTs, respectively. Both the $^{Nat}$DWCNTs and $^{Iso}$DWCNTs were oxidized in air at 520°C for 4h to open the ends of the tubes and to introduce defects into the walls of the nanotubes. Finally, the as-prepared and the oxidized DWCNTs were annealed together at 1500°C for 1h in a dynamic vacuum (<$10^{-6}$ mbar) to synthesize the confined carbyne, which were marked as the one-step annealed and stepwise annealed samples, respectively.

*Raman Characterization*: Raman spectroscopy was carried out with a Raman spectrometer (LabRAM HR800, Horiba Jobin-Yvon) equipped with a liquid nitrogen cooled CCD detector, a 50x objective (numerical aperture: 0.5), a 1000 μm pin-hole, a 100 μm slit, an 800 mm confocal length, and a 600 gr/mm grating. Several laser excitations with wavelengths of 488, 514, 532, 568 and 633 nm were applied to achieve resonance. The laser spot size is of about 2 μm and the spectral resolution is about 2 cm$^{-1}$. The measurements were taken at ambient conditions using a laser power below 0.5 mW to avoid heating effect (no shift of G-band was observed compared to the G-band measured at much lower power). All the spectra at different laser excitation energies were calibrated against the spectral lines of a Neon lamp. For easy comparison, all the Raman spectra were normalized to the G-band intensity.

## Acknowledgements

This work was supported by the University of Vienna and Austrian Science Fund (FWF, P27769-N20). L.S. acknowledges the financial support from the National Natural Science Foundation of China (51902353), Guangdong Basic and Applied Basic Research Foundation (No. 2019A1515011227) and the State Key Laboratory of Optoelectronic Materials and Technologies (No. OEMT-2021-PZ-02 and No. OEMT-2022-ZRC-01). F.S. was supported by the Ministry of Innovation and Technology and the National Research, Development and Innovation Office (NKFIH) of Hungary within the Quantum Information National Laboratory and the grant K137852.

**Keywords:** isotope labelling • confined carbyne • Raman spectroscopy • atom exchange • nanotube reorganization